\documentclass[aps,prl,twocolumn,groupedaddress]{revtex4}
\begin{document}


\title{Any is not all: EPR and the Einstein-Tolman-Podolsky paper}


\author{Charles Tresser}
\email[]{charlestresser@yahoo.com}
\email[]{ PACS: 03.65.Ta}
\affiliation{IBM, P.O. Box 218, Yorktown Heights, NY 10598, U.S.A.}


\date{\today}

\begin{abstract}
 In Bohm's version of the EPR \emph{gedanken} experiment, the spin of the second particle along any vector is minus the spin of the other particle along the same vector.  It seems that either the choice of vector along which one projects the spin of the first particle influences at superluminal speed the state of the second particle, or naive realism holds true (\emph{i.e.,} the projections of the spin of any EPR particle along all the vectors are determined before any measurement occurs).  Naive realism is negated by Bell's theory that originated and is still most often presented as related to non-locality, a relation whose necessity has recently been proven to be false.  I advocate here that the solution of the apparent paradox lies in the fact that the spin of the second particle is determined \emph{along any vector}, but \emph{not along all vectors}. Such an any-all distinction was already present in quantum mechanics, for instance in the fact that the spin can be measured along any vector but not at once along all vectors, as a result of the Uncertainty Principle.  The time symmetry of the any-all distinction defended here is in fact reminiscent of (and I claim, due to) the time symmetry of the Uncertainty Principle described by Einstein, Tolman, and Podolsky in 1931, in a paper entitled ``Knowledge of Past and Future in Quantum Mechanics" that is enough to negate naive realism and to hint at the any-all distinction.  A simple classical model is next built, which captures aspects of the any-all distinction: the goal is of course not to have a classical exact model, but to provide a caricature that might help some people.
\end{abstract}

\pacs{03.65.Ta}

\maketitle

%
%
%
%
The stated goal of the Einstein-Podolsky-Rosen (or \emph{EPR}) paper \cite{EPR} was to prove the \emph{incompleteness} of Quantum Mechanics (or \emph{QM}). That paper used \emph{entangled pairs of particles}, \emph{i.e.,} two-particles states whose wave functions cannot be written as tensor products.  Following now Einstein's own views (see, \textit{e.g.,} \cite{EinsteinIdeasAndOpinions}, \cite{Schilpp}, as well as \cite{FineShaky} and references therein) rather than \cite{EPR}, what is measured on one particle, say among two pre-selected Conjugate Observables (or \emph{COs}), influences what has to be the wave function for the second particle (which can be computed using Wave Packet Reduction), even if the two measurements are space-like separated.  Hence one has to accept either the incomplete character of QM since the wave function is assumed to be the most complete description of a state, or physics to be \emph{non-local} (\emph{i.e.,} to permit instant propagation of some effects of some causes).  The COs used in EPR were the position and momentum.  Instead of using the quite general configuration of \cite{EPR}, we will consider entangled pairs of spin-$\frac{1}{2}$ particles that are prepared, following  Bohm \cite{Bohm}, in the so-called \emph{singlet state} that is rotation invariant (see, \textit{e.g.,} \cite{Bohm}, p. 400) and given along any vector by:
\[
\Psi(x_1,x_2)=\frac{1}{\sqrt{2}}(| +\rangle _1\otimes| -\rangle_2-| -\rangle_1\otimes|
+\rangle_2)\,,
\]
as far as the description of the spin part of the wave function of the singlet is concerned.  Then the COs are chosen to be the projections of the spin on two orthogonal vectors that are also orthogonal to the axis along which the particles separate. The singlet  state is invariant by rotation, and the + and - signs that appear in the formula for $\Psi(x_1,x_2)$ correspond to the projections of the spins of the two particles along the same arbitrary direction.  In particular, the singlet state is such that along any vector, the total projected spin is 0 (there is a corresponding case for light polarization that is more practical for experiments to date but that will not be used here \cite{BohmAharonov1957}). If one measures the normalized spin of a particle of the pair, say particle 1, along some vector $\vec{a}$, and the result $s_1(\vec{a})$ is +1 or -1, then by Wave Packet Reduction (WPR), the new state becomes respectively  $\Psi'(x_1,x_2)= | +\rangle _{1,\vec{a}}\otimes| -\rangle_{2,\vec{a}}\,,
$ or $\Psi'(x_1,x_2)=| -\rangle_{1,\vec{a}}\otimes|
+\rangle_{2,\vec{a}}\,.
$
The version of EPR using singlets is called \emph{EPRB} and one also speaks of \emph{EPRB pairs}.  Bell later proposed \cite{Bell}  to use arbitrary pairs of vectors to project the spins of the two particles of EPRB pairs.  Notice that once the projection of the spin of particle 1 is measured along $\vec{a}$ so that $s_1(\vec{a})=\varepsilon\in \{ +1,\, -1\}$, by using the WPR  one gets the second particle in a known spin state, as if it was prepared in that state: $s_2(\vec{a})=-\varepsilon$.  Knowing this spin projection of the second particle along $\vec{a}$, QM tells us that the probability to find  $s_2(\vec{b})=s_2(\vec{a})$ is $P(\theta)=\frac {1+\cos(\theta)}{2}$ with $\theta$ the angle between $\vec{a}$ and $\vec{b}$ (this is Malus' law for polarization, but restated here for spin-$\frac{1}{2}$ particles).  Hence the probability $P(\theta)$ for the spin of the second particle along $\vec{b}$ to be the same that the spin of the first one along $\vec{a}$ (\emph{i.e.,} the probability to get $s_2(\vec{b})=s_1(\vec{a}))$ is $P'(\theta)=1-P(\theta)=\frac {1-\cos(\theta)}{2}$. The relation between readings along  $\vec{a}$ for the first particle and along $\vec{b}$ for the second particle can also be formulated, as Bell did in \cite{Bell}, in terms of the correlation between the observations along these vectors, \emph{i.e.,} the average value $<\cdot>$ of the product of $s_1(\vec{a})$
by $s_2(\vec{b})$: one then gets $<s_1(\vec{a})\cdot s_2(\vec{b})>=-\vec{a} \cdot \vec{b}$, as Bell noticed in \cite{Bell} with no further comment.  More importantly, Bell also proved a theorem (based on a inequality that relates several correlations of spin projections along pairs of vectors) in \cite{Bell}, under two main assumptions:

- A1) There are \emph{predictive Hidden Variables} (HVs)  (where predictive means that all variables, including the observables, are determined once all variables are taken into account, even if no one can know what the observables are determined to be),
an hypothesis later sometimes replaced by assuming \emph{classical realism} (according to which, \emph{e.g.,} the projections of the spin of any EPR particle along all directions are determined, even if out of our reach, before measurement).  The statistics of QM augmented by HVs or by classical realism is anyway assumed to be the same as for usual QM.

- A2) \emph{Locality} to the effect that nothing travels faster than light, so that the spin of one particle of an EPRB pair does not depend on the setting of the measurement tool used for the other one, at least when one assumes that the measurements (or other evaluations of observables as they may occur under assumption A1) on the particles are spatially separated. 

\medskip
I now present an \emph{over-simplified} and thus \emph{false} version of the Bell reasoning, where instead of assuming HVs or classical realism, one assumes that several measurements can be made on any of the particles of an EPRB pair.  One can then measure the probability $P(\theta)$.  Notice that $P'(\theta)$ can also be indirectly measured since the spin $s_{\vec{c}}(e)$ of the electron $e$ along $\vec{c}$ is minus the spin $s_{\vec{c}}(p)$ of the positron $p$ along the same vector $\vec{c}$, when $(e,p)$ is an EPRB pair.  As was recalled above, QM teaches us that: 
\[
(\circ)\qquad P'(\theta)=1-P'(\theta)=\frac {1-\cos(\theta)}{2}\,,
\] 
which has been confirmed by many experiments (see, \emph{e.g.,} \cite{Aspect1999} for a review).

\medskip
Consider now 4 vectors $\vec{a}$, $\vec{a'}$, $\vec{b}$, and $\vec{b'}$, and set:

\noindent
$\theta _1=(\vec{a},\vec{b})$, $\theta _2=(\vec{b},\vec{a'})$, and $\theta _3=(\vec{a'},\vec{b'})$.

\medskip
Because the spin along a vector is a binary data, we must have for instance:
\[
(*)\qquad P'(\theta _1)+P'(\theta _2)+P'(\theta _3)\geq P'(\theta _1+\theta _2+\theta _3 ) \,.
\]
Specializing (*) to $\theta _1=\theta _2=\theta _3=\frac{\pi}{4}$ so that $\theta _1+\theta _2+\theta _3=\frac{3\cdot\pi}{4}$, we see that $P'(\theta _1)$, $P'(\theta _2)$ and $P'(\theta _3)$ are all equal to $\frac{2-\sqrt 2}{4}$ (about $.15$), while $P'(\theta _1+\theta _2+\theta _3)=\frac{2+\sqrt 2}{4}$ (about $.85$), so that $(*)$ would reduce to the false inequality $1>\sqrt 2$.  Under the same assumption, one also gets:
\[
(**)\qquad P'(\theta _1)+ P'(\theta _2)+P(\theta _1 +\theta _2)\geq 1\,,
\]
which provides the same false inequality $1>\sqrt 2$ when specialized to $\theta _1=\theta _2=\frac{\pi}{4}$.  Inequalities such as $(*)$ or $(**)$ come from \emph{gedanken} experiments where one measures several projections of the spins of the particles: 2 projections on each particle for $(*)$ and 2 on one side and 1 on the other side for $(**)$: see  \cite{Tresser2005_1} and references therein.  While this discussion, and the fact in particular that $(\circ)$ has been experimentally verified, should have bothered a ninetieth century physicist, the 1927 Uncertainty Principle \cite{Heisenberg1927} implies in particular that \emph{while the spin measurement for $e$ or $p$ in any given EPRB pair could be made along {\bf any} vector, such measurement cannot be done along {\bf all} vectors}, and in fact not along more than one vector at once on each side so that $(*)$ and even $(**)$ need too many measurements.  

Thus the Uncertainty Principle prevents the \emph{gedanken} experiments that lead to inequalities such as $(*)$ or $(**)$ from becoming actual experiments by violation of a law of physics (we say that we are dealing with \emph{counter-natural} experiments). Thus the impossible inequalities that one gets are the consequence of the counter-natural character of the \emph{gedanken} experiments leading to inequalities such as $(*)$ or $(**)$ in the setting that we have expounded where many measurements are needed.  Any projection measurement is possible, but not all at once because of the Uncertainty Principle, a principle so well accepted that no one would look for further explanation of the false inequalities that we have obtained.

\smallskip
The \emph{``any-all"} distinction was not spelled out in the previous papers that I know about, and although this was not needed for what was attempted in \cite{Tresser2005_1} or \cite{Tresser2005_2}, I have come to realize the essential nature of the distinction between the elements of the \emph{(any, all)} pair, where :

\smallskip
\noindent
- \emph{``any"} means that the spin of an EPRB particle is determined along ``any one vector that one would choose",

\smallskip
\noindent
 - \emph{``all"} means that the spin of an EPRB particle is determined along ``all vectors".
 
\smallskip
\noindent
\textbf{Remark.} \emph{There is a trivial physical impossibility in the use of ``all", which is linked to the fact that, for instance, ``all vectors" contains infinitely many vectors: we will thus understand ``all vectors" as ``as many vectors as one wants", so that in particular classical quantities can be measured along ``all vectors" (in this weakened sense) in clear contrast with QM.}

\smallskip
The price of overs-simplification was to hurt the reasoning by bumping head-on into the Uncertainty Principle. In the actual Bell setting, one can consider directly $P'(\theta)$ as well as $P(\theta)$, and one gives sense to $P(\theta)$ and/or $P'(\theta)$ for several values of the angle $\theta$ by augmenting QM by predictive HVS or by classical realism (assumption A1).  Any inequality that one gets in the previous oversimplified setting can as well be obtained in the actual Bell setting, and this holds true for instance for  $(*)$ or $(**)$, but with the new hypotheses, these inequalities become instances of \emph{Bell Inequalities} (BI). 

Now, the 1931 paper by Einstein, Tolman, and Podolsky (hereafter ETP) \cite{ETP}, tells us that the Uncertainty Principle works as well when going back in time.  This causes $P(\theta)$ or $P'(\theta)$  to make sense \emph{for {\bf any} $\theta$ but not for {\bf all} $\theta$'s at once on any given EPRB pair}, and the BIs such as
 $(*)$ or $(**)$ are just as counter-natural in view of \cite{ETP} as they were under the former multiple measurements hypotheses because of \cite{Heisenberg1927}.  While the projection of the spin of an EPRB particle along any vector is determined assuming realism, if said realism is to not contradict the (backward) Uncertainty Principle, such projections can only make sense for at most one vector per particle of each EPRB pair and in particular not for all vectors.  Yet, because of the endemic any-all confusion, one is often erroneously faced with the alternative: 

\smallskip
\noindent
\emph{``non-locality of QM" \emph{vs} ``the spin of  $\mathbf{p}$ is determined before measurement along all vectors". }

\smallskip
The magic of Bell's inequality, and of Bell's Theorem (according to which QM augmented by predictive HVs or by classical realism leads to contradiction when assuming locality) has made lots of people forget about the deep counter-natural character of the setting.  Whether or not people mention he augmentation of QM, it is most often locality which is charged with the impossible inequalities that one reaches, and not counter-naturality. Noticeable exception to collective blindness are the 1972 paper by de la Pe{\~n}a,  Cetto, and Brody \cite{PenaCettoBrody1972} (summarized on p. 312 of  \cite{Jammer1974}) and work by Fine from the early 1980'th \cite{Fine1982a}, \cite{Fine1982b} (see also  \cite{Tresser2005_1} and  \cite{Tresser2005_2} on such matters).

Here are some highlights of the understanding that one can get now, in view of \cite {ETP}:

- 1) Bell's theorem allows one to disqualify \emph{naive realism} (and \emph{naive HVs theories}, but it should be enough to discuss naive realism since it is implied by naive HVs), where \emph{``naive"} stands for \emph{incompatible with QM and in particular with the Uncertainty Principle and its corollaries}.  

- 2) The 1931 paper \cite{ETP} by Einstein, Tolman, and Podolsky allows one to disqualify the same type of naive realism; in some sense  \cite{ETP} is precisely what makes such realism ``naive" (recall that the Uncertainty Principle was only 4 years old in 1931 and that many distinguished authors seem to not have appreciated the counter-natural character of the setting needed to prove a Bell's type theorem till now!).

- 3) The ETP paper \cite{ETP} might be a clue to understand why Einstein avoided using counter-natural discourses when dealing with the issue of the completeness of QM.  Whether that is the case remains a mystery to me, especially in view of the fact that Podolsky (and Rosen) fell in the trap (as so many did much later).

Anyway:
 
- 4) Because of the counter natural character of the setting for Bell's type theorems, such theorems cannot be verified experimentally, nor can it be verified experimentally that QM violates a Bell's inequality.

- 5) One cannot establish a Bell type inequality which makes sense if one uses an augmentation of QM that involves a HV theory or a form of realism that permits at most one measurement and/or one value among a set of COs on any particle. 

- 6) One cannot establish a Bell type inequality which makes sense if one uses an augmentation of QM that involves a HV theory or a form of realism that permits at most one measurement and/or one value among a set of conjugate observable on any particle, and possibly two values if one can get information on one particle using another particle that is entangled to the first one.

- 7) Bell's theorem and \cite{ETP} cannot disqualify non-naive realism or HVs, but I do not know of any good reason to expect an augmentation of QM by non-naive HVs or realism to work, nor to expect the contrary (I personally consider that the hopes for a completion of QM are a last form of anthropocentrism since very small scale escape space-time geometry; see \cite{Tresser2005_2} on that matter).  

\smallskip
I have provided in \cite{Tresser2005_1}  a setting for a Bell's type theorem (leading to $(**)$) that is simple enough to allow to better see the fact that it is the counter-natural character that is a problem, so that Okam's razor should cut out non-locality (see also\cite{Tresser2005_2}).  Recall that non-locality cannot be disposed of otherwise than by using Okam's razor or something of that kind because it is not susceptible of experimental evidence nor of counter evidence.

\smallskip
Bell's Theorems allow one to show the incompatibility with QM of the type of realism that one can extract from the original EPR paper \cite{EPR} written by Podolsky.  This is the same naive form of realism that one can see also in a text written 50 years later by Rosen in \cite{Rosen1985} (in both cases, the naive realism is hidden in a subtle way) but it is quite different (see  \cite{Tresser2005_2} and references therein) from the form of realism used by Einstein.  Expositions of the non-completeness of QM by Einstein himself avoid counterfactuals: see, \emph{e.g.,} the 1936 text reproduced in \cite{EinsteinIdeasAndOpinions} and the later piece in \cite{Schilpp} where Einstein also stays away from naive realism.
In effect, Bell type theorems, and further entanglements analysis such as for the so-called GHZ states, have been used a lot to disqualify realism \cite{GHZ1989}, \cite{Mermin1990GHZ3}, \cite{GHSZ1990GHZ3}, \cite{Tresser2005_2}.  However it is often by restricting the analysis to naive realism (without pointing out its naive character), hence restricting to realism of a sort that Einstein by himself has always avoided (at least as soon as the 1930's) but still attributing such realism to Einstein, making the attacks worthwhile (a brutal -even if not desired- form of Einstein bashing, as follows from the rest of this paper in view of \cite{Jammer1974} and \cite{FineShaky}). 

\bigskip
I next proceed to construct a model that may help one picture what is going on in the any-all business (nothing more than a naive image is to be expected from classical models, as far as I understand; some may not want any such false images because they are of no help given the profound difference between the classical and the quantum world. Anyway, only vague hints can be expected from the classical models that I propose here for the any-all discussion.
 
To this modest end, I begin with a model with even more modest goals.  We consider the $z$-axis, or vertical line $\Delta _z$ (thus defined by $x=y=0$) in $\mathbf{R} ^3$, and the vertical tube $\mathcal{T}$ with section the square $\sup(|x|, |y|)=1.1$.  Each face of the tube is equipped with (separation) lines with constant $z$ that are equally spaced, at distance 1 of each other and we assume that the space in between two lines is colored, alternatively in black (or marked ``1")  and white (or marked ``-1").  Now the line closest to the plane $z=0$ is different on each face, with a \emph{shift} from $z=0$ given by:
 
\noindent
- $z=Z_1=0$ for the first face of $\mathcal{T}$, $x=1.1$, 

\noindent
- $z=Z_2=\frac{1}{4}$ for the second face of $\mathcal{T}$, $y=1.1$, 
 
\noindent 
- $z=Z_3=\frac{1}{2}$ for the third face of $\mathcal{T}$, $x=-1.1$, 

\noindent
- $z=Z_4=\frac{3}{4}$ for the fourth face of $\mathcal{T}$, $y=-1.1$.

Imagine now a ball $B_\alpha$ of diameter 1 centered on the point $z=\alpha$ of $\Delta _z$, where $\alpha$ is any real number that is not of the form $Z_i+n$ with $i\in \{1,2,3,4\}$ and $n\in \mathbf{Z}$.  Imagine that the ball $B_\alpha$, when pushed quasi-horizontally to the $k^{\rm th}$ face $\mathcal{F}_k$ of  $\mathcal{T}$, has to fall with its center horizontally aligned with the center $C(k,\alpha )$ of the unique colored square of $\mathcal{F}_k$ that contains the horizontal line $z=\alpha$. After being so pushed, the \emph{$k$-value} of $B_\alpha$ is $1$ or $-1$ depending on the color of the square of  $\mathcal{F}_k$ with center $C(k,\alpha )$.  Before being so ``measured" against a face of $\mathcal{T}$, any of the \emph{$k$-values} of the ball $B_\alpha$ is determined, but while not all of them can be measured at once, all of them are indeed determined.  Let us try to improve a bit the status, keeping in mind that the classical and quantum Worlds are definitely different.

\smallskip
In a modification of the above setting, the ball $B'_\beta$ is oval of vertical length equal to 2, and by pushing to a face, one gets to the $C(k,\beta )$ determined, not by the geometric center of $B'_\beta$, but by its field dependent  center of gravity $G(F_k,B'_\beta )$ as it gets revealed by the field $F_k$  in place near the face $\mathcal{F}_k$ of $\mathcal{T}$.  The way the dust that fils  $B'_\beta$ gets concentrated to $G(F_k,B'_\beta )$ under the influence of the field $F_k$ is \emph{deterministic but quite complex, sensitive upon initial conditions, and irreversible}.  
The determination is made by experimentation (pushing to a face): one cannot use a (discrete) machine computation as any knowledge must be gained in finite time and there is sensitivity upon initial conditions. The rare cases leeading to ambiguity are eliminated.
Thus, before being ``measured" against a face of $\mathcal{T}$, any of the $k$-values of the ball $B'_\beta$ is determined as the measured value if the $k$ value gets to be measured, but it is not true that all of the $k$-values are determined: the impossibility of joined valuation is exactly as strong as the impossibility of simultaneous measurement.  In fact, the only access to determination is measurement, although all that determines the position of $G(F_k,B'_\beta )$ in $B'_\beta $ is there so that there is no room for indeterminacy, but only an effective impossibility for the determination: the goal was not to have a classical exact model, but a model allowing to accept what may elude our full understanding forever (as advocated in \cite{Tresser2005_2}).

\smallskip
The reader will easily check that the tube structure can be made richer, so that all of the shifts happen instead of just four of them: then one finds also a correlation similar to Malus' law (correlation = $\vec{a}\cdot\vec{b}$) between the values of successive measures along $\vec{a}$ and then $\vec{b}$.

\end{document}